\documentclass[12pt]{article} 
\usepackage{ucs} 
\usepackage[utf8x]{inputenc} 
\usepackage[T1]{fontenc} 
\usepackage{amsmath,amssymb} 
\usepackage{graphicx} 
\usepackage[unicode]{hyperref} 
\hypersetup{bookmarksopen=true, bookmarksnumbered=true, pdfstartview={FitH}, pdfborder={0 0 0}, colorlinks=true, linkcolor=red, linktocpage=true, citecolor=blue, urlcolor=cyan, unicode=true, pdftitle={Darboux Coordinates and Instanton Corrections in Projective Superspace}, pdfauthor={PMCrichigno, DJain}, pdfsubject={Instanton Corrections to HyperKähler Moduli Space Metrics}, pdfkeywords={N=2 Projective Superspace, }{Hypermultiplets, }{Hyperkähler Manifolds, }{Instantons}}
\PrerenderUnicode{ä}\PrerenderUnicode{×}\PrerenderUnicode{¹}\PrerenderUnicode{²}\PrerenderUnicode{³}
\usepackage{geometry} 
\usepackage{fancyhdr} 
\usepackage[nottoc,notlof,notlot]{tocbibind} 
\usepackage[titles]{tocloft} 

\usepackage{parskip} 
\newcommand\references[1] {\begin{thebibliography}{[99]} #1 \end{thebibliography}}
\numberwithin{equation}{section} 
\newcommand\cen[1] {\begin{center}#1\end{center}}
\newcommand\equ[1] {\begin{equation}#1\end{equation}}

\newcommand\eqs[1] {\begin{align}#1\end{align}}

\renewcommand\( {\left(}
\renewcommand\) {\right)}
\newcommand\F {{\cal F}}
\renewcommand\G {{\cal G}}
\renewcommand\L {{\cal L}}
\newcommand\N {{\cal N}}
\renewcommand\O {{\cal O}}

\newcommand\V {{\cal V}}

\DeclareUnicodeCharacter{"0393}{\Gamma}
\DeclareUnicodeCharacter{"0394}{\Delta}
\DeclareUnicodeCharacter{"0398}{\Theta}
\DeclareUnicodeCharacter{"039B}{\Lambda}
\DeclareUnicodeCharacter{"039E}{\Xi}
\DeclareUnicodeCharacter{"03A0}{\Pi}
\DeclareUnicodeCharacter{"03A3}{\Sigma}
\DeclareUnicodeCharacter{"03A5}{\Upsilon}
\DeclareUnicodeCharacter{"03A6}{\Phi}
\DeclareUnicodeCharacter{"03A8}{\Psi}
\DeclareUnicodeCharacter{"03A9}{\Omega}
\DeclareUnicodeCharacter{"03B1}{\alpha}
\DeclareUnicodeCharacter{"03B2}{\beta}
\DeclareUnicodeCharacter{"03B3}{\gamma}
\DeclareUnicodeCharacter{"03B4}{\delta}
\DeclareUnicodeCharacter{"03B5}{\epsilon}
\DeclareUnicodeCharacter{"03B6}{\zeta}
\DeclareUnicodeCharacter{"03B7}{\eta}
\DeclareUnicodeCharacter{"03B8}{\theta}
\DeclareUnicodeCharacter{"03D1}{\vartheta}
\DeclareUnicodeCharacter{"03B9}{\iota}
\DeclareUnicodeCharacter{"03BA}{\kappa}
\DeclareUnicodeCharacter{"03BB}{\lambda}
\DeclareUnicodeCharacter{"03BC}{\mu}
\DeclareUnicodeCharacter{"03BD}{\nu}
\DeclareUnicodeCharacter{"03BE}{\xi}
\DeclareUnicodeCharacter{"03C0}{\pi}
\DeclareUnicodeCharacter{"03C1}{\rho}
\DeclareUnicodeCharacter{"03C3}{\sigma} 
\DeclareUnicodeCharacter{"03C4}{\tau}
\DeclareUnicodeCharacter{"03C5}{\upsilon}
\DeclareUnicodeCharacter{"03C6}{\phi}
\DeclareUnicodeCharacter{"03D5}{\varphi}
\DeclareUnicodeCharacter{"03C7}{\chi}
\DeclareUnicodeCharacter{"03C8}{\psi}
\DeclareUnicodeCharacter{"03C9}{\omega}
\DeclareUnicodeCharacter{"21D0}{\Leftarrow}
\DeclareUnicodeCharacter{"0212B}{\AA}
\DeclareUnicodeCharacter{"266A}{\eighthnote}
\DeclareUnicodeCharacter{"266B}{\twonotes}

\newcommand\arXivid[1] {\href{http://arxiv.org/abs/#1}{\tt arXiv:#1}} 
\newcommand\cmp[3] {{\it Commun.\ Math.\ Phys.\ } \href{http://inspirehep.net/search?ln=en&ln=en&p=find+j+"Commun.Math.Phys.,#1,#3"&of=hb&action_search=Search&sf=&so=d&rm=&rg=25&sc=0}{{\bf #1} (#2) #3}} 

\newcommand\jhep[3]{{\it JHEP\ } \href{http://inspirehep.net/search?ln=en&ln=en&p=find+j+"JHEP,#1,#3"&of=hb&action_search=Search&sf=&so=d&rm=&rg=25&sc=0}{{\bf #1} (#2) #3}}
\newcommand\npb[3] {{\it Nucl.\ Phys.\ }{\bf B #1} (#2) #3}
\newcommand\pl[4] {{\it Phys.\ Lett.\ }{\bf #1 #2} (#3) #4}
\newcommand\prl[3] {{\it Phys.\ Rev.\ Lett.\ } \href{http://inspirehep.net/search?ln=en&ln=en&p=find+j+"Phys.Rev.Lett.,#1,#3"&of=hb&action_search=Search&sf=&so=d&rm=&rg=25&sc=0}{{\bf #1} (#2) #3}}

\begin{document}
\pagenumbering{alph}
\title{\Huge Darboux  Coordinates and Instanton Corrections in Projective Superspace}
\author{P. Marcos Crichigno\footnote{\href{mailto:crichigno@insti.physics.sunysb.edu}{crichigno@insti.physics.sunysb.edu}} , Dharmesh Jain\footnote{\href{mailto:djain@insti.physics.sunysb.edu}{djain@insti.physics.sunysb.edu}}\bigskip\\ \emph{C. N. Yang Institute for Theoretical Physics}\\ \emph{State University of New York, Stony Brook, NY 11790-3840}}
\date{} 
\maketitle
\thispagestyle{fancy}
\rhead{YITP-SB-12-09} 
\lhead{\today}
\begin{abstract}
\normalsize By demanding consistency of the Legendre transform construction of hyperk\"{a}hler metrics in projective superspace, we derive the  expression for the Darboux coordinates on the hyperkähler manifold. We apply these results to study the Coulomb branch moduli space of 4D, $\N=2$ super-Yang-Mills theory (SYM) on $\mathbb R^3 \times S^1$, recovering the results by GMN. We also apply this method  to study the electric corrections to the moduli space of 5D, $\N=1$ SYM on $\mathbb R^3 \times T^2$ and give  the Darboux coordinates explicitly.  
\end{abstract}

\newpage
\pagenumbering{Roman}
\cfoot{\thepage}\rhead{}\lhead{}
\tableofcontents

\newpage
\pagenumbering{arabic}
\section{Introduction}
Since the work of Seiberg and Witten \cite{Seiberg:1994rs, Seiberg:1994aj}, the structure of  $\N=2$ theories in four dimensions has been extensively explored, leading to important insights into the dynamics of gauge theories. A recent area of progress in this field is the study of the Coulomb branch moduli space of $\N=2$ theories on $\mathbb R^3 \times S^1$, first analyzed in \cite{Seiberg:1996nz}. It has received renewed attention due its relation to the Kontsevich-Soibelman (KS) wall-crossing formula \cite{KS} for $\N=2$ theories in the work by Gaiotto, Moore, and Neitzke (GMN) \cite{GMN}. As described by GMN, the KS formula ensures the continuity of the metric on the moduli space. Alternatively, demanding continuity of the metric provides a physical proof of the wall-crossing formula. The central idea in \cite{GMN} was to find an efficient description of the moduli space metric and its corrections due to BPS instantons. Such a description was given in terms of the holomorphic Darboux coordinates $(\eta_e,\eta_m)$ by making crucial use of twistor techniques. One of the important results of their work is that the magnetic coordinate $\eta_m$ is  given in terms of the electric coordinate $\eta_e$ (in the mutually local case)  by
\equ{\eta_{m}=\eta_m ^{sf}+ \frac{i}{2} \int_{l_{+}} \frac{d \zeta'}{2 \pi i \zeta'}\frac{\zeta+\zeta'}{\zeta-\zeta'}  \ln{(1- e^{i  \eta_e})}-\frac{i}{2} \int_{l_{-}} \frac{d \zeta'}{2 \pi i \zeta'}\frac{\zeta+\zeta'}{\zeta-\zeta'}  \ln{(1- e^{-i  \eta_e})}\,. \label{GMN0}}

One of the goals of this paper is to use the formalism of $\N=2$ Projective Superspace to present a simple derivation and generalization of this formula for any hyperk\"{a}hler manifold described by  $\O(2p)$ multiplets. Our analysis is based on the projective Legendre transform, which dualizes the $\O(2p)$ supermultiplet $\eta_e$ to an ``arctic'' supermultiplet $\Upsilon$.  We will see that $\eta_m$ is the imaginary part of $\zeta^{p-1}\Upsilon$ and is given by
\equ{\eta_m= \frac{i}{2} \oint_{C_0} \frac{d \zeta'}{2 \pi i \zeta'}\frac{1}{\zeta-\zeta'} \left[ \zeta \(\frac{\zeta}{\zeta'}\)^{p-1}+  \zeta' \(\frac{\zeta'}{\zeta}\)^{p-1} \right] \frac{\partial f}{\partial \eta'_e}\,, \label{Main formula0}}
where $f$ is  the projective Lagrangian describing the manifold and $C_0$ is a contour around the origin. Such expression can also be obtained from gluing conditions for the Darboux coordinates, as done in  \cite{Alexandrov et al}  (see \cite{Alexandrov:2011va} for a recent review and references therein).  Our derivation, however, is based on requiring the consistency of the Legendre transform by imposing the condition that $\Upsilon$ is regular at $\zeta=0$. The kernel in  (\ref{Main formula0})  is understood as a  projector ensuring this consistency condition. 
 
In the specific case of the periodic Taub-NUT metric, by Poisson resummation of the usual projective Lagrangian describing it, we recover (\ref{GMN0}). A natural generalization of $f$ to incorporate mutually nonlocal corrections leads to the integral TBA equation. We then consider the doubly-periodic Taub-NUT metric, deriving an expression for the Darboux coordinates, which reduces to  (\ref{GMN0}) in a particular limit. This metric was recently studied in \cite{Haghighat:2011xx} to  describe the electric corrections to the moduli space of five-dimensional SYM on $\mathbb R^3 \times T^2$.
  
This paper is organized as follows: In Section \ref{Preliminaries}, we review some basic elements of projective superspace and then derive (\ref{Main formula0}) by demanding consistency of the Legendre transform construction of hyperk\"{a}hler manifolds in Section \ref{Darboux Coordinates}. Then, we apply this result to the moduli space of $\N=2$ SYM on $\mathbb R^3 \times S^1$, recovering the results by GMN in Section \ref{Section 4d}. In Section \ref{Section 5d}, we study the moduli space of  $\N=1$ SYM on $\mathbb R^3 \times T^2$ and conclude in Section \ref{Conclusions} with a summary and discussion of open problems.

\section{Preliminaries}\label{Preliminaries}
In this section, we review some elements of $\N=2$ projective superspace \cite{ULMR} and the construction of hyperk\"{a}hler metrics \cite{Hitchin:1986ea}. A recent review of essential aspects of the relation between projective superspace and hyperkähler manifolds can be found in \cite{Lindstrom:2008gs}.

\subsection{Projective Superspace}
The algebra of $d=4$, $\N=2$  supercovariant derivatives is
\equ{ \{D_{i \alpha},D_{j \beta}\}=0\,, \qquad  \{D_{i \alpha}, \bar{D}^j_{\dot \beta}, \}=i \, \delta^j_i \partial_{\alpha \dot \beta} \,, }
where $i, j=1,2$ are $SU(2)_R$ indices and $\alpha, \dot \alpha$ are spinor indices. Projective superspace is defined as the Abelian subspace parametrized by a coordinate $\zeta\,∈\,\mathbb{CP}^1$ and spanned by the combinations
\equ{\nabla_{\alpha}(\zeta)= D_{2 \alpha}+\zeta D_{1 \alpha}\,, \qquad \bar{\nabla}_{\dot \alpha}(\zeta)=\bar{D}_{\dot \alpha}^1-\zeta \bar{D}_{\dot \alpha}^2\,,}
where $ D_{1 \alpha}$ and $\bar{D}_{\dot \alpha}^1$ are $\N=1$ derivatives and $ D_{2 \alpha}$ and $\bar{D}_{\dot \alpha}^2$ are the generators of the extra supersymmetry. These combinations satisfy
\equ{\{∇_{\alpha},∇_{\beta}\}=\{∇_{\alpha},\bar{∇}_{\dot \beta}\}=0\,.}
Projective superfields are then defined to satisfy the constraints:
\equ{\nabla_{\alpha}\Upsilon=\bar{\nabla}_{\dot \alpha}\Upsilon=0\,. \label{projective constraints}}
There are several types of projective supermultiplets, characterized by their $\zeta$-dependence.  We shall mainly focus on two: real $\O(2p)$ and (ant)arctic supermultiplets. The first class of multiplets are polynomial in $\zeta$, with its powers ranging from $-p$ to $p$, and real under the \textit{bar} conjugation (complex conjugation composed with the antipodal map: $ζ→-1/ζ$). In particular, the $\O(2)$ multiplet is defined by
\equ{\eta_e=\frac{a}{\zeta}+\theta_e - \bar{a} \zeta\,.}
 It follows from (\ref{projective constraints}) that $a$ and $\theta_e$ are $\N=1$ chiral and real linear superfields, respectively. The second class of multiplets are arctic and antarctic superfields, which are defined to be analytic around the north pole ($\zeta=0$) and south pole ($\zeta=\infty$), respectively, $i.e.$,
\equ{\Upsilon=\sum_{n=0}^{\infty} \Upsilon_n \zeta^n\, ,  \qquad \bar{\Upsilon} = \sum_{n=0}^{\infty}\bar{\Upsilon}_{n}\left(\frac{-1}{\zeta}\right)^n \,. }
From (\ref{projective constraints}), it follows that only the two lowest components of the arctic superfield are constrained $\N=1$  superfields (chiral and complex linear, respectively), while the remaining (infinite) components are auxiliary (unconstrained) complex superfields.

\subsection{Hyperkähler Manifolds}\label{Hyperk\"{a}hler manifolds}
Here we review the construction of  hyperk\"{a}hler metrics in projective superspace \cite{Hitchin:1986ea, Lindstrom:2008gs, Lindstrom:1987ks, Ivanov:1995cy}. Given an arbitrary analytic function $f(\eta_e;\zeta)$, one defines the function
\equ{F(a,\bar a, \theta_e)\equiv \oint_C \frac{d \zeta}{2 \pi i \zeta} f(\eta_e; \zeta)\,,\label{ProjLag}}
where $C$ is an appropriately chosen contour, which typically depends on the choice of $f$ (referred to as the projective Lagrangian henceforth). The Legendre transform of $F$ serves as the K\"{a}hler potential $K$ for a hyperk\"{a}hler manifold, $i.e.$,
\equ{K(a,\bar a, v+ \bar v) = F(a, \bar a, \theta_e) - (v+\bar v)\, \theta_e\,, \qquad F_{\theta_e}=v+\bar v\,, \label{Legendre Transform components}}
where $v$ is an $\N=1$ chiral superfield. Note that K\"{a}hler metrics described in this way automatically have an isometry, associated to shifts of $\text{Im}(v)$. The resulting metric is of the Gibbons-Hawking form
\equ{ds^2=\frac{1}{V(x)} \(d \theta_m +  A\)^2 +V(x) d\vec x \cdot d\vec x \,,}
where $a=x^1+ i x^2, \,\theta_e= x^3$ and $d V = ⋆d A$, with
\eqs{V=\oint_{C} \frac{d \zeta'}{2 \pi i \zeta'} \frac{\partial^2 f}{\partial \eta'^2_e}\,, \qquad A=\frac{1}{2}∮_{C}\frac{dζ'}{2πiζ'}\(\frac{1}{ζ'}da+ζ'd\bar{a}\)\frac{∂^2f}{∂η'^2_e}\,.\label{ALEmet}}
An important class of metrics are $A_{N-1}$ ALE metrics and can be described in this way by taking
\equ{f(\eta_e)=\sum_{k}\left(\eta_e-\eta_k\right) \log \left(\eta_e-\eta_k\right)\,,\label{ALE_N}}
where $\eta_k$ are constant $\O(2)$ multiplets simply giving the position $\vec x_k$ of $N$ mass points.  For this Lagrangian, the contour in (\ref{ProjLag}) is an 8-shaped contour $\tilde C$ enclosing the two roots of $\eta_e-\eta_k=0$. Indeed, using (\ref{ALE_N}) in (\ref{ALEmet}) gives the harmonic function
\equ{V=  \sum_{k} \oint_{\tilde{C}} \frac{d \zeta}{2 \pi i \zeta} \frac{1}{\eta_e-\eta_k}= 2 \sum_{k} \frac{1}{|\vec x-\vec x_k| } \,\label{formV}}
and the corresponding $A$. Taking an infinite superposition of mass points along $\theta_e$, $i.e.$, taking $\eta_k=k$ and $N\rightarrow \infty$, the metric becomes periodic  along this direction\footnote{Strictly speaking, $V$ is logarithmically divergent and must be properly regularized. It should be understood that this has been done in what follows.}. This metric (commonly referred to as the Ooguri-Vafa metric) was discussed by  Ooguri and Vafa in \cite{Ooguri:1996me} and Seiberg and Shenker in \cite{Seiberg:1996ns}. Following the terminology of \cite{Gaiotto:2011tf}, we will refer to it as the periodic Taub-NUT (PTN) metric. We will refer to a PTN metric which is periodic along two directions as the doubly-periodic Taub-NUT  (dPTN) metric.

A hyperk\"{a}hler manifold has three K\"{a}hler forms $\omega^{(2,0)}, \omega^{(1,1)}$ and $ \omega^{(0,2)}$, which can be conveniently organized into 
\equ{\varpi=\omega^{(2,0)}+\omega^{(1,1)} \zeta -\omega^{(0,2)} \zeta^2\,.}
This combination can be further written as:
\equ{\varpi= i \zeta\,d \eta_e \wedge d \eta_m\,, \label{Symplectic Form GH}}
with
\eqs{d \eta_e= \frac{da}{\zeta}+d \theta_e - \zeta d\bar{a}\,, \qquad d \eta_m = d \theta_m + i A + \frac{i V}{2}\(\frac{1}{\zeta} da + \zeta d \bar a\)\, . \label{d magnetic GH}}
Since the symplectic form $\varpi$ takes the canonical form in (\ref{Symplectic Form GH}), the set $(\eta_e, \eta_m)$ are referred to as Darboux coordinates. The main purpose of the coming sections is to find an explicit expression for $\eta_m$ in terms of $f(\eta_e;\zeta)$.

\subsection{Duality and Symplectic Form}\label{Duality and symplectic form}
One can alternatively describe these hyperk\"{a}hler manifolds in terms of an arctic superfield $\Upsilon$, rather than in terms of an $\mathcal{O}(2)$, by a duality relating these two multiplets  \cite{Ivanov:1995cy, GonzalezRey:1997qh}.  In terms of $\N=1$ components, this is based on the Legendre transform (\ref{Legendre Transform components}) exchanging a real linear superfield by a chiral superfield. It is similarly described in terms of projective superfields as follows: One relaxes the condition of $\eta_e$ being an $\O(2)$ multiplet, imposing this through a Lagrange multiplier $\Upsilon +\bar \Upsilon$. Integrating out $\Upsilon$ leads to the original description in terms of $\eta_e$, while integrating out $\eta_e$ leads to a dual description in terms of $\Upsilon$. That is, one defines
\equ{\tilde f(\Upsilon+\bar \Upsilon;\zeta)=f(\eta_e; \zeta) -(\Upsilon +\bar \Upsilon) \eta_e\,,}
with the standard Legendre transform relations
\equ{\frac{\partial f}{\partial \eta_e} =\Upsilon +\bar{\Upsilon}\,, \qquad  \frac{\partial  \tilde f}{\partial \Upsilon}=- \eta_e  \label{real part upsilon}\,.}
The main advantage of this setup (for our purposes)  is that one can define a holomorphic symplectic two-form  \cite{Lindstrom:2008gs} that captures the essential aspects of the hyperk\"{a}hler geometry (see also \cite{Kuzenko:2009wr} for related results). 
This is based on the observation that arctic superfields have infinitely many unconstrained $\N=1$ fields $\Upsilon_i$, for $i\geq 2$, which must be integrated out. These equations of motion imply that
\equ{\tilde \Upsilon \equiv \zeta \frac{\partial  \tilde f}{\partial \Upsilon}=-\zeta \eta_e\, \label{Upsilon tilde}}
is also an arctic superfield. Thus, one can define a 2-form $\varpi$ by 
\equ{\varpi=d\Upsilon \wedge d \tilde \Upsilon=\omega^{(2,0)}+\omega^{(1,1)} \zeta -\omega^{(0,2)} \zeta^2 \label{Symplectic form upsilon} \,.}
In other words, $\Upsilon$ and $\tilde \Upsilon$ are (by construction) Darboux coordinates for the holomorphic symplectic form $\varpi$. Note that they are regular at $\zeta=0$, while $(\bar \Upsilon,\bar{\tilde \Upsilon})$ are regular at $\zeta=\infty$, and
\equ{\varpi=-\zeta^2 \overline{\varpi}= -\zeta^2 d\bar{\Upsilon} \wedge d \bar{\tilde{\Upsilon}}.}
Thus, up to the twisting factor $\zeta^2$, there is a symplectomorphism relating north pole and south pole coordinates and the generating function is precisely $\tilde f(Υ+\bar \Upsilon)$, giving a geometric interpretation to the $\N=2$ projective Lagrangian.

\section{Darboux Coordinates}\label{Darboux Coordinates}
As seen in Section \ref{Duality and symplectic form}, the projective Legendre transform provides an  expression for a set of Darboux coordinates, namely $(\Upsilon, \tilde \Upsilon)$. The coordinate $\tilde \Upsilon$ is given by (\ref{Upsilon tilde}) whereas only the real part (under bar conjugation) of $\Upsilon$ is determined by (\ref{real part upsilon}), $i.e.$
\equ{\Upsilon= \frac{1}{2} \frac{\partial f}{\partial \eta_e} + i \eta_m\,,\label{Upsilon}}
where we have introduced $\eta_m=\bar \eta_m$ as the (undetermined) imaginary part of $\Upsilon$. The crucial observation \cite{MR Notes} is that $\Upsilon$ is actually completely determined by a consistency requirement on the whole construction. Recall that the constraint of $\eta_e$ being an $\O(2)$ multiplet was imposed through a Lagrange multiplier, assuming that $\Upsilon$ was an arctic superfield. However, the first term on the $\text{r.h.s.}\,$ of (\ref{Upsilon}) contains negative powers of $\zeta$ and therefore, the consistency requirement is that these should be canceled by $\eta_m$. This $\eta_m$  is precisely the magnetic coordinate we are after, since we find from (\ref{Upsilon tilde}) and (\ref{Upsilon}) that
\equ{\varpi=d\Upsilon \wedge d \tilde \Upsilon= i \zeta\,d \eta_e \wedge d \eta_m\,
\label{Symplectic form}}
coincides with (\ref{Symplectic Form GH}). To determine $\eta_m$, we introduce the antarctic projector
\equ{\Pi_N \equiv \oint_{C_0} \frac{d \zeta'}{2 \pi i }\frac{1}{\zeta-\zeta'}\,, \qquad  \Pi_N^2=\Pi_N\,,  \qquad \Pi_N \bar \Pi_N =0\,, \label{Arctic Projector}}
where $C_0$ is a closed contour around the origin (see Appendix \ref{A:proj}). This projector annihilates the non-negative powers of $\zeta$.
Thus, the consistency requirement is simply
\equ{\Pi_N \Upsilon=0\,. \label{Consistency conditions}}
\newpage
It is easy to see that
\equ{\eta_m= \theta_m+\Big( i \Pi_N   - i \bar \Pi_N\Big)\frac{1}{2}\frac{\partial f}{\partial \eta_e}\,,\label{Solution eta O(2)}}
with $\theta_m=\bar \theta_m$, solves the consistency condition\footnote{Indeed, from (\ref{Upsilon}), (\ref{Solution eta O(2)}), and using the properties in (\ref{Arctic Projector}), we see that
\equ{\Pi_N \Upsilon=  \Pi_N\(\frac{1}{2} \frac{\partial f}{\partial \eta_e}+ i\left[\theta_m+ \Big(  i \Pi_N   - i \bar \Pi_N\Big)\frac{1}{2} \frac{\partial f}{\partial \eta_e}\right]\)= \Pi_N\(\frac{1}{2} \frac{\partial f}{\partial \eta_e}\)- \Pi_N\(\frac{1}{2} \frac{\partial f}{\partial \eta_e}\)=0\,.\nonumber}}. 
We can rewrite (\ref{Solution eta O(2)}) in a more familiar form. From (\ref{Arctic Projector}), we see that the projectors combine into
\equ{ i  \Pi_N   - i \bar \Pi_N  =  i \oint_{C_0} \frac{d \zeta'}{2 \pi i \zeta'}\frac{\zeta+\zeta'}{\zeta-\zeta'} }
and hence
\equ{\eta_m=\theta_m+\frac{i}{2} \oint_{C_0} \frac{d \zeta'}{2 \pi i \zeta'}\frac{\zeta+\zeta'}{\zeta-\zeta'} \frac{\partial f}{\partial \eta'_e}\,,
\label{main formula}}
recovering the expression obtained in \cite{Alexandrov et al}. The derivation of this expression,  by ensuring and making manifest  that $\Upsilon$ is arctic, is one of the main results of this  paper. This condition is enforced by the projector $(\zeta+\zeta')(\zeta-\zeta')^{-1}$ and will be extended below to include $\O(2p)$ multiplets.  

We can easily check that from (\ref{main formula}) we recover the expression (\ref{d magnetic GH}) for Gibbons-Hawking metrics. Acting with $d$ on $\eta_m$, we have
\eqs{d \eta_m&=  dθ_m+\frac{i}{2} \oint_{C_0} \frac{d \zeta'}{2 \pi i \zeta'}\frac{\zeta+\zeta'}{\zeta-\zeta'} \frac{\partial^2 f}{\partial \eta_{e}^{'2}}\(d \eta'_e-d \eta_e\)  \nonumber \\ 
                         &= dθ_m+\frac{i}{2} \oint_{C_0} \frac{d \zeta'}{2 \pi i \zeta'}\left[ \(\frac{1}{\zeta}+\frac{1}{\zeta'}\)da+(\zeta+\zeta')d \bar a \right] \frac{\partial^2 f}{\partial \eta^{'2}_e} \nonumber \\
                        &=dθ_m+i A+\frac{i V}{2}\(\frac{1}{ζ}da+ζ\,d\bar{a}\)\,.}
In the first line, we have added a term proportional to $d\eta_e$, which gives no contribution to the symplectic form (\ref{Symplectic form}). In the last line, we have used the definitions (\ref{ALEmet}), assuming that the contour giving the K\"{a}hler potential is $C_0$. 

Although the derivation of (\ref{main formula}) requires a contour enclosing only a singularity at the origin, note that choosing the contour to be the one defining the K\"{a}hler potential gives the correct symplectic form.  This expression provides a systematic way of constructing Darboux coordinates for any hyperk\"{a}hler manifold described by  an $\mathcal{O}(2)$ multiplet $\eta_e$ and projective Lagrangian $f$. We will use this in the following sections to describe instanton corrections to moduli spaces of SYM theories.  \vspace{30pt}

\subsubsection*{Semiflat Geometry and the $c$-map\footnote{This section is based on \cite{MR Notes}.}}
It is clear from (\ref{main formula}) that, unlike $\eta_e$, the magnetic coordinate $\eta_m$  will not be an $\O(2)$ in general, this depending on the singularity structure of $f(η_e;\zeta)$. A special case however is when the rigid $c$-map \cite{CFG:cmap, Gates:1999zv, RVV:cmap} (see Appendix \ref{cmap}) can be applied.  According to the $c$-map, 
\equ{f^{sf}(η_e;\zeta)=-i\(\frac{\F\(\zeta \eta_e\)}{\zeta^2}   - \overline{\F}\(-\frac{\eta_e}{\zeta}\)  \,\zeta^2\)\,,\label{c-map}}
where $\F(W)$ is the $\N=2$ holomorphic prepotential. The $c$-map gives the contribution from na\"{\i}ve dimensional reduction, without taking into account the effect of BPS particles. Thus, one expects $\eta_m$ to be given by an $\O(2)$. However, by the direct substitution of (\ref{c-map}) in (\ref{main formula}), we see that this is not the case. This is resolved by recalling that the Darboux coordinates are defined up to terms that vanish in the symplectic form.  In fact, we can add such a term to the definition of $\eta_m$ that does lead to an $\O(2)$, namely
\eqs{\Upsilon&= \frac{1}{2} \frac{\partial f^{sf}}{\partial \eta_e} + i \(\eta_m^{sf}  -\frac{1}{2}\( \frac{\F'}{\zeta}-\bar \F' \zeta\)  \)\\
 &= -\frac{i \F'(\zeta \eta_e)}{\zeta} +i \eta_m^{sf}\,.\label{Upsilon sf}}
From the fact that $\F'(\zeta \eta_e)=\F'(a+\theta_e \zeta - \bar a \zeta^2)$ is regular at the origin, the condition that $\Upsilon$ in (\ref{Upsilon sf}) is arctic is simply solved by 
\equ{\eta_m^{sf}=\frac{\F'(a)}{\zeta}+\theta_m - \bar{\F}'(\bar a) \, \zeta\,.\label{MSF}}
Therefore, na\"{\i}ve electric-magnetic duality $a\rightarrow a_D =\F'(a)$ holds. In general,  dyonic multiplets have the form $η_{γ}^{sf}=\frac{Z_γ}{\zeta}+\theta_γ - \bar Z_γ\,\zeta\,,$ where the central charge is $Z_γ=n_e a+n_m a_D$ with $n_e$ and $n_m$ being the electric and magnetic charges, respectively. Once BPS instanton corrections are included,  the magnetic coordinate is no longer an $\O(2)$ since the total Lagrangian is
\equ{f=f^{sf}+f^{inst}\,,\nonumber}
where $f^{inst}$ is \textit{not} of the form (\ref{c-map}). Thus, the full magnetic coordinate is  given in general  by 
 \equ{\eta_m=\eta_m^{sf}+\frac{i}{2} \oint_{C_0} \frac{d \zeta'}{2 \pi i \zeta'}\frac{\zeta+\zeta'}{\zeta-\zeta'} \frac{\partial f^{inst}}{\partial \eta'_e}\,, \label{eta_m sf plus inst}}
where $\eta^{sf}_m$  is given by (\ref{MSF}).

\subsubsection*{Generalization to $\mathcal{O}(2p)$ Multiplets} 
Our construction so far includes only hyperk\"{a}hler manifolds described by $\O(2)$ multiplets, but it can be easily extended to the case of $\O(2p)$ multiplets by a generalization of the Legendre transform relating $\Upsilon$ to an $\O(2p)$ multiplet $\eta_e$ \cite{GonzalezRey:1997qh} . Additional factors of $\zeta$ have to be introduced in the Legendre transform to impose the corresponding constraint on $\eta_e$,  namely 
\equ{\tilde f=f -\(\zeta^{p-1} \Upsilon +\(-\zeta\)^{-(p-1)} \bar \Upsilon\) \eta_e }
with the relations
\equ{\frac{\partial f}{\partial \eta_e} =\zeta^{p-1} \Upsilon +\(-\zeta\)^{-(p-1)} \bar \Upsilon\,, \qquad \tilde \Upsilon ≡ \zeta \frac{\partial  \tilde f}{\partial \Upsilon}=-\zeta^{p} \eta_e\,. \label{Upsilon O2pA}}
Thus, we now have
\equ{\zeta^{p-1} \Upsilon=  \frac{1}{2} \frac{\partial f}{\partial \eta_e} + i \eta_m\,, \label{def mag coord O2p}}
and the symplectic form is still given  by
\equ{\varpi=d\Upsilon \wedge d \tilde \Upsilon= i \zeta \,d \eta_e \wedge d \eta_m\,.  \nonumber}
The magnetic coordinate $\eta_m$ will again be determined by the requirement that the resulting superfield $\Upsilon$ is arctic.  From (\ref{Upsilon O2pA}) it follows that $\frac{\partial f}{\partial \eta_e}$ contains powers  $\zeta^{n}$ with $|n|\geq(p-1)$ only. Thus, $\eta_m$ in (\ref{def mag coord O2p}) is required to cancel the powers $\zeta^n$ with $n< -(p-1)$ of $\frac{\partial f}{\partial \eta_e}$ and we cannot add a $\zeta$-independent term, contrary to the $\O(2)$ case. Using the corresponding projectors, we then find
\equ{\eta_m= \frac{i}{2} \oint_{C_0} \frac{d \zeta'}{2 \pi i \zeta'}\frac{1}{\zeta-\zeta'} \left[ \zeta \(\frac{\zeta}{\zeta'}\)^{p-1}+  \zeta' \(\frac{\zeta'}{\zeta}\)^{p-1} \right] \frac{\partial f}{\partial \eta'_e}\,. \label{eta_m O2p}}
The corresponding semiflat contribution can be determined using  the $c$-map prescription for $\O(2p)$ multiplets given in \cite{Gates:1999zv}.

A metric which is described, for example,  by an $\mathcal{O}(4)$ multiplet is the Atiyah-Hitchin metric, characterizing the moduli space of two monopoles and the moduli space of three-dimensional SYM.  It would be interesting to compare (\ref{eta_m O2p}) to the Darboux coordinates given in \cite{AHcoords}.  In the remainder of the paper, we will restrict ourselves to $\O(2)$ multiplets and apply our results to the study of moduli spaces of pure SYM theories with eight supercharges in $d=4$ and $d=5$.

\section{\texorpdfstring{$\N=2$ SYM on $\mathbb{R}³×S¹$}{N=2 SYM on R³×S¹}}\label{Section 4d}
In this section, we apply our construction to the study of the Coulomb branch of pure $\N=2$ SYM with gauge group $SU(2)$, first analyzed in \cite{Seiberg:1996nz}.  The bosonic content of the four-dimensional theory consists of a complex scalar field $a$ and a gauge field $A_μ$. Upon dimensional reduction on a circle $S^1$ of radius $R$ (which we set to 1 in this section), the gauge field decomposes as $A_{\mu}\rightarrow (A_i,A_4)$, giving a three-dimensional photon and a real scalar field. Since in three dimensions the photon itself is dual to a scalar field,  the moduli space of supersymmetric vacua is four-dimensional. Furthermore, due to the amount of supersymmetry it is hyperk\"{a}hler.  It can be parameterized by the vev of the vector multiplet scalar field, $a$, in addition to the gauge-invariant electric and magnetic Wilson loops\footnote{We have normalized the angular variables $\theta_{e,m}$ to have period 1.} 
 \equ{\theta_e \equiv  \frac{1}{2π}\oint_{S^1_4} A_4 \,, \qquad \theta_m \equiv  \frac{1}{2π}\oint_{S^1_4} A_{D,4}\,.}
Naïve dimensional reduction of the 4D SYM action results in a  3D sigma model with a target space metric of Gibbons-Hawking form, specified by the ``semiflat'' potential $V^{sf}=\text{Im}τ$, where $τ$ is the usual complexified 4D gauge coupling. However, the BPS particles from the four-dimensional theory can wrap the compactification circle $S^1$, generating instanton corrections to the semiflat metric in the compactified theory, which we discuss next.

\subsection{Mutually Local Corrections}
Following \cite{GMN}, we begin by assuming that all the BPS particles are mutually local and choose a duality frame in which there are no magnetically charged particles. This leads to a shift isometry in $\theta_m$ and therefore the space is naturally described by the $\O(2)$ multiplet
\equ{\eta_e = \frac{a}{\zeta}+\theta_e -\bar a \zeta \,.}
Integrating out a hypermultiplet of electric charge $q$ (which we set to 1 here) leads to a Taub-NUT metric. Summing over the infinite tower of Kaluza-Klein momenta $k$ along the $S^1$ turns it into the periodic Taub-NUT metric described in Section \ref{Hyperk\"{a}hler manifolds}.  Thus, the projective Lagrangian is given by 
\equ{f(\eta_e)=\sum_{k=-\infty}^{\infty} (\eta_e-k) \log (\eta_e-k)\,.}
Recall that here each term in the Lagrangian is to be integrated along an 8-figure contour around the roots of $\eta_e-k=0$. To isolate instanton contributions, we perform a Poisson resummation, yielding\footnote{Poisson resummation works as follows:
\equ{\sum_{n=-\infty}^{\infty} f(n)= \sum_{k=-\infty}^{\infty} \hat{f}(k)\,,\qquad \hat{f}(k)=\int_{-\infty}^{\infty} dx \, e^{-2 \pi i k x} f(x)\,. \nonumber}
In (\ref{4d inst potential}) we have omitted the divergent $n=0$ term.}
\eqs{f&=f^{sf}+f^{inst}\,; \nonumber \\ f^{sf}&=-i\(η_e^2\log\(\frac{ζη_e}{Λ}\)-η_e^2\log\(\frac{-η_e}{ζ\bar{Λ}}\)\)\,, \label{4d sf potential} \\
f^{inst}&=i \, s \sum_{n>0} \frac{1}{n^2}e^{i n \eta_e} \theta(s)+i \,  s \sum_{n<0} \frac{1}{n^2}e^{i n \eta_e} \theta(-s)\,,\label{4d inst potential}}
where $Λ$ is the UV cutoff and $s\equiv \text{sign}\left[\text{Im}\(\eta_e\)\right]$. The semiflat Lagrangian $f^{sf}$ has been included using the $c$-map prescription described previously, with the 1-loop prepotential $\F(W)\sim W^2 \log W^2$. The full magnetic coordinate is then given by (\ref{eta_m sf plus inst}). Note that since the Heaviside functions $\theta(\pm s)$ in $f^{inst}$ contain $\zeta$, they restrict the integration contour. Using the identity
\equ{\text{Im}\(\eta_e\) = (1+|\zeta|^2)\text{Im}\left(\frac{a}{\zeta}\right)\,,}
we see that $θ(\pm s)$ imposes the BPS ray condition\footnote{Our conventions in the definition of $\eta_e$ differ by a factor $i$ with those of GMN, and so does the definition of the BPS rays.} $l_{\pm}=\left\{ \zeta: \text{sign}\left[\text{Im}\(\frac{a}{\zeta}\)\right]=\pm1\right\}$, leading to
\eqs{\oint_{C_0} f^{inst}(η_e)=i\int_{l_+} \mathrm{Li}_2\(e^{i \eta_e}\)-i\int_{l_-} \mathrm{Li}_2\(e^{-i \eta_e}\)\,, \label{f^inst}}
where we have used the series expansion for $\mathrm{Li}_2(x) = ∑_{n=1}^{∞} \frac{x^n}{n^2}$. Substituting (\ref{f^inst}) in (\ref{eta_m sf plus inst}) finally gives
\equ{\eta_{m}=\eta^{sf}_m+ \frac{i}{2} \int_{l_{+}} \frac{d \zeta'}{2 \pi i \zeta'}\frac{\zeta+\zeta'}{\zeta-\zeta'}  \ln{(1- e^{i \eta_e})}-\frac{i}{2} \int_{l_{-}} \frac{d \zeta'}{2 \pi i \zeta'}\frac{\zeta+\zeta'}{\zeta-\zeta'}  \ln{(1- e^{-i \eta_e})}\,, \label{GMN main formula 2}}
where $\eta_m^{sf}$ is given by (\ref{MSF}). Thus, we have recovered GMN's result for the mutually local case. We now discuss the mutually nonlocal case.

\subsection{Mutually Nonlocal Corrections}\label{Mutually Nonlocal Corrections}
Inspired by the analytic and asymptotic properties of (\ref{GMN main formula 2}), an integral equation (of the form of a TBA equation) for the Darboux coordinates in the mutually nonlocal case was derived in \cite{GMN}. The natural proposal to include dyonic multiplets is that  each BPS particle of charge $\gamma$ contributes independently to the projective instanton Lagrangian, with a weight  given by the multiplicity of each state $\Omega(\gamma';u)$, $i.e.$,
\equ{f^{inst}= i \sum_{\gamma'}  \Omega(\gamma' ;u) \, \mathrm{Li}_2 \left(\sigma(\gamma') e^{i \eta_{\gamma'}}\right)\, \theta(s_{γ'})\,. \label{f inst non-local}}
Here $\gamma=\(n_e,n_m\)$ is a  vector in the two-dimensional charge lattice with the antisymmetric product $\langle \gamma, \gamma' \rangle = n_e n_m'- n_e' n_m$, $\sigma(\gamma)=(-1)^{n_e\,n_m}$,  $\eta_{\gamma}= n_e \eta_e +n_m \eta_m$, and  $s_{γ} =\text{sign}\left[\text{Im}\(\frac{Z_{\gamma}}{\zeta}\)\right]$ that defines the BPS ray $l_{γ}$. From (\ref{eta_m sf plus inst}), it is natural to write the following integral equation for the dyonic coordinate
\eqs{ \eta_{\gamma}=\eta_{\gamma}^{sf}+\frac{i}{2} \sum_{\gamma'} \langle \gamma', \gamma \rangle   \oint_{C_0} \frac{d \zeta'}{2 \pi i \zeta'}\frac{\zeta+\zeta'}{\zeta-\zeta'}  \frac{\partial f^{inst}}{\partial \eta'_{\gamma'}}\,. \label{TBA general}}
Inserting (\ref{f inst non-local}) above leads to
\equ{\eta_{\gamma}= \eta_{\gamma}^{sf} + \frac{i}{2}\sum_{\gamma'}\Omega(\gamma' ;u) \langle \gamma', \gamma \rangle  \int_{l_{γ'}} \frac{d \zeta'}{2 \pi i \zeta'}\frac{\zeta+\zeta'}{\zeta-\zeta'}  \ln \left( 1-\sigma(\gamma') e^{i \eta'_{\gamma'}}\right)\,, }
corresponding to the TBA equation that determines the exact moduli space metric. Note  that the Darboux coordinates played the central role in the analysis by GMN, being in some sense the fundamental objects. In the current setting, the fundamental object (which behaves additively and contains all the geometric information) is the projective Lagrangian $f$. The Darboux coordinates are determined by it through the integral equation (\ref{TBA general}).

\section{\texorpdfstring{$\N=1$ SYM on $\mathbb{R}³×T²$}{N=1 SYM on R³×T²}}\label{Section 5d}
Minimally supersymmetric Yang-Mills in five dimensions  has an interesting BPS spectrum, containing not only electrically charged particles, but also magnetically charged strings  and dyonic instantons \cite{Lambert:1999ua}. Since the theory is non-renormalizable by power-counting it should be viewed as a field theory with a cutoff. In this sense, one can still ask what are the quantum corrections to the moduli space. This was first studied in \cite{Seiberg:1996bd}, where the exact Coulomb branch metric was determined. More recently, the compactification of this theory on $T^2$ was studied in \cite{Haghighat:2011xx}, giving an important first step in analyzing the Coulomb branch metric of the compactified theory. Since dimensional reduction of this theory to four dimensions leads to the theory discussed in the previous section, compactification of the five-dimensional theory on $T^2$ gives a (two-parameter) generalization of the moduli space studied above. 

The bosonic content of this theory consists of a real scalar $\sigma$ and the gauge field $A_{\hat \mu}$. Upon dimensional reduction to three dimensions, the gauge field reduces according to $A_{\hat \mu}\rightarrow (A_{i},A_4,A_5)$, leading again to a four-dimensional moduli space. The two electric coordinates $\varphi_1, \varphi_2$ and the ``magnetic'' coordinate $\lambda$ are defined by  
 \equ{\varphi_1 \equiv  \frac{1}{2π} \oint_{S^1_4} A_4 \, ,\qquad \varphi_2 \equiv  \frac{1}{2π} \ \oint_{S^1_5} A_5\,, \qquad \lambda \equiv \int_{T^2} B\,,}
where $B_{\hat \mu \hat \nu}$ is the (2-form) dual of the photon $A_{\hat \mu}$. Under large gauge transformations, these variables are periodic and parameterize a torus $T^2$. Due to the electric particles running around these two compactified dimensions, the Coulomb branch metric inherits the modular properties of the torus and has an isometry in $\lambda$. A full analysis of the moduli space must include the effect of dyonic instantons, as well as the mutually nonlocal effect of  magnetic strings wrapping the whole $T^2$, which will break the isometry in $\lambda$. In this paper, we focus only on the projective description of the electric corrections to the moduli space metric, hoping that this will help in incorporating the effect of magnetic strings as well. 

\subsection{Electric Corrections}\label{Mutually local corrections 5d}
Here we apply the methods of Section \ref{Darboux Coordinates}  to find the corrections to $\eta_m$, due to electric particles running along  the two compact directions. It is clear that the metric in this case is simply the dPTN metric. For simplicity, we discuss first the projective description of this metric in the case of a rectangular torus and then for a generic torus with complex structure $\tau$.
 
\subsubsection*{Rectangular Torus}
Consider a rectangular torus with radii $R_1,R_2$ and complex structure $\tau=i\frac{R_1}{R_2}$. We  define the doubly periodic $\O(2)$ multiplet by
\equ{\eta_e=\frac{\sigma  R_2+i \varphi_2}{2 R_2\zeta}+\frac{\varphi_1}{R_1} -\frac{(\sigma  R_2-i  \varphi_2)}{2 R_2}\zeta\,.}
With this definition, the projective Lagrangian $f$ for the dPTN metric has the form (\ref{ALE_N}) with
\equ{\eta_k=  \frac{1}{R_1}k_1 +\frac{i}{2 R_2} \(\frac{1}{\zeta}+ \zeta\) k_2\ \equiv a_1 k_1 +a_2 k_2 \,.}
For convenience, rather than concentrating on the calculation of $f$, in this section we will focus on the Gibbons-Hawking potential $V$, given by
\equ{V= \sum_{\vec{k}}  \oint_{\tilde{C}} \frac{d \zeta}{2 \pi i \zeta}\,\frac{1}{\eta_e-a_1 k_1 -a_2 k_2}\,. \label{V5dRec}}
As before, $\tilde{C}$ is an $8-$shaped contour enclosing the poles of the integrand for each $\vec k$, leading to a doubly periodic Gibbons-Hawking potential. This potential is linearly divergent and as in the PTN case should be understood to be properly regularized.  We now perform a double Poisson resummation. Resumming over $k_1$ first gives
\eqs{V&=V^{(0)}+V^{(1)}\,,\nonumber \\
V^{(0)}&= -R_1 \oint_{C_0}  \frac{d \zeta}{2 \pi i \zeta}  \sum_{k_2} \log{\left[ \zeta R_1 \(\eta_e - a_2 k_2 \) \right]}+c.c.\,,  \label{First Poisson resummed0}
 \\
V^{(1)}&=-i R_1 ∮_{C_0} \frac{d \zeta}{2 \pi i \zeta} \sum_{k_2} \sum_{n_1 \neq 0} e^{i n_1 R_1\(\eta_e - a_2 k_2\)} s\,\theta\(n_1 s\)\,,  \label{First Poisson resummed1}
}
where $s=\text{sign}\left[\text{Im}\( \eta_e - a_2 k_2 \)\right]$. Here $V^{(0)}$  is a superposition of shifted semiflat potentials of Section \ref{Section 4d}. We now show that it leads to the effective gauge coupling $1/g_4(a)^2$ due to the dimensional reduction from 5D to 4D  \cite{Lawrence:1997jr}, and it reduces (after Poisson resummation) to the semiflat potential in the $R_2\rightarrow 0$ limit. Performing the integral around the origin  in (\ref{First Poisson resummed0}) gives
\equ{V^{(0)}= -R_1 \sum_{k_2} \log{\(   \frac{\sigma  R_2+i (\varphi_2-k_2)}{2 R_2}  \)}+c.c. 
 =R_1\sum_{n_2 \neq 0} \frac{1}{|n_2|} e^{-(R_2 |n_2 \sigma|- i n_2 \varphi_2)}\,, \label{V0 PR}}
where  we performed a Poisson resummation for the second equality. This in fact matches the result in \cite{Lawrence:1997jr} (see also \cite{Haghighat:2011xx}). In the four-dimensional limit, 
\equ{V^{(0)} \xrightarrow{R_2 \rightarrow 0} V^{sf}_{4D}=R_1 \( \log a + \log \bar a\)\,,}
 where $a =\frac{\sigma  R_2+i \varphi_2}{2 R_2}$, which coincides with  the potential derived from (\ref{4d sf potential}). The contribution to the magnetic coordinate is given by
\equ{\eta_m^{(0)}=\frac{\F'(a)}{\zeta}+\frac{\lambda}{R_1} - \bar{\F}'(\bar a) \, \zeta\,, \qquad \F(a) =\frac{1}{4 R_2^2}\left[\text{Li}_3\(e^{2 a R_2}\)\theta(-\sigma)+\text{Li}_3\(e^{-2 a R_2}\)\theta(\sigma)\right]  \,, \label{SF 5d}}
where  we have integrated  (\ref{V0 PR}) twice with respect to $a$ to determine $\F(a)$. 

Now we turn to $V^{(1)}$, which in the $R_2 \rightarrow 0$ limit reduces to the instanton corrections in the four-dimensional theory. The contour in (\ref{First Poisson resummed1}) splits into two rays $l_{\pm}$, and integration along these rays ensures that the limit $R_2 \rightarrow 0$ is well defined. In fact, in this limit the sum over $k_2$ is localized at $k_2=0$, $i.e.$,
\equ{V^{(1)}  \xrightarrow{R_2 \rightarrow 0} V^{inst}_{4D}=-i R_1 ∮_{C_0} \frac{d \zeta}{2 \pi i \zeta} \sum_{n_1 \neq 0} e^{i n_1 R_1\eta_e} s\,\theta\(n_1 s\),}
which is the Gibbons-Hawking potential one would get from (\ref{4d inst potential}). (One should rescale $a \rightarrow  R_1 a$ in the four-dimensional case for comparison.)

For finite $R_2$, Poisson resumming (\ref{First Poisson resummed1}) leads to\footnote{Here we have dropped a term in the exponent 
\equ{e^{i n_1 R_1\eta_e +i\,\text{Im}\(\eta_e\)\frac{2 R_1 R_2|ζ|^2}{\(1+|ζ|^2\)\text{Re}(ζ)}\left[\frac{n_2}{R_1}-\frac{i n_1}{2R_2}\(\frac{1}{\zeta}+\zeta\)\right]}\,, \nonumber} because we choose the contour enclosing the origin along which $\text{Im}(\eta_e)=\pm \epsilon$. In the limit $\epsilon \rightarrow 0$ this term does not contribute to the integral, which becomes simply an integral around the origin.}
\equ{V^{(1)}= -\oint_{C_0} \frac{d \zeta}{2 \pi i \zeta} \sum_{\substack{n_1 \neq 0\\n_2  \in \mathbb Z}} \frac{e^{\frac{i n_1 \eta_e}{a_1}}}{  a_2 n_1 - a_1 n_2   }\,. \label{Contour int 5d Rectangular1}}
Note that after the double Poisson resummation, the contour in (\ref{Contour int 5d Rectangular1}) remains a closed contour, enclosing only the essential singularity at the origin (and not the simple poles). By residue integration, we find
\equ{V^{(1)}=R_1R_2 \sum_{\substack{n_1 \neq 0\\n_2 \in \mathbb Z}} \frac{1}{\sqrt{n_1^2 R_1^2+n_2^2 R_2^2}} e^{i\(n_1\varphi_1+n_2\varphi_2\)-\left|\sigma\right| \sqrt{n_1^2 R_1^2+n_2^2 R_2^2} } \,.}
Combining the two contributions, we have
\equ{V=V^{(0)}+V^{(1)}=R_1R_2\sum_{\vec{n} \in \mathbb Z^{2'}} \frac{1}{\sqrt{n_1^2 R_1^2+n_2^2 R_2^2}} e^{i\(n_1\varphi_1+n_2\varphi_2\)-\left|\sigma\right| \sqrt{n_1^2 R_1^2+n_2^2 R_2^2} }\,,}
which matches the expression for $U_{1-loop}$ in \cite{Haghighat:2011xx}. 
Integrating twice with respect to $\eta_e$ (and dropping a possible linear term, which does not contribute to $\eta_m$), we find
\equ{f^{(1)}=∑_{\substack{n_1 \neq 0\\n_2∈\mathbb{Z}}}\frac{a_1^2}{n_1^2\(n_2a_1-n_1a_2\)}\,e^{\frac{i n_1\eta_e}{a_1}}\,. \label{f1}}
As explained in \cite{Haghighat:2011xx}, the corrections due to $f^{(1)}$ to the Coulomb branch metric should coincide with the corrections to the hypermultiplet moduli space due to D1 instantons in type IIB theory. Indeed, we find that the projective Lagrangian $f^{(1)}$ matches with that given in \cite{5D:LSTV}. Now, putting all the elements together, the magnetic coordinate for the dPTN metric  finally reads
\equ{\eta_{m}=\eta_m^{(0)} +\frac{i}{2}\oint_{C_0} \frac{d \zeta'}{2 \pi i \zeta'}\,\frac{\zeta+\zeta'}{\zeta-\zeta'}\,  \frac{\partial f^{(1)} }{\partial \eta_e'}  \,. \label{etam5d}}
In summary, the magnetic coordinate  contains two parts: the $\eta_m^{(0)}$ part from the na\"{\i}ve 5D to 4D reduction, which becomes $\eta_m^{sf}$ in the 4D limit, and the remaining part, which reduces to $\eta_{m}^{inst}$.

\subsubsection*{Generic Torus} 
To consider a generic torus with complex structure $\tau$, we perform a modular transformation from the rectangular case. Under the $SL(2,\mathbb Z)$ symmetry group of the torus, the complex structure $\tau=\tau_1+i \tau_2$ and the coordinates transform as
\equ{\tau \rightarrow \frac{a \tau +b}{c \tau +d}\,, \qquad \left(
\begin{array}{c}
\varphi_2 \\ 
\varphi_1
\end{array}\right) \rightarrow  \left(
\begin{array}{cc}
a &  c\\ 
b &  d
\end{array}\right)\left(
\begin{array}{c}
\varphi_2 \\ 
\varphi_1 
\end{array}\right)\,,}
with $ad-bc=1$. 
The electric coordinate for a generic torus then becomes\footnote{For a generic torus with complex structure $\tau=\tau_1 +i \tau_2$, we perform the $SL(2,\mathbb Z)$ transformation $\varphi \rightarrow M^T  \varphi$, where $M$ is given by
\equ{M=\frac{1}{\sqrt{\tau_2}}\left(
\begin{array}{cc}
τ_2 & \tau_1\\ 
0 & 1
\end{array}\right)\,.\nonumber} 
The metric  transforms according to $g\rightarrow (M^{-1})^T g\, M^{-1}$,  leading to (\ref{general metric}).}
\equ{\eta_e=\frac{1}{2\zeta}\(\sigma +i\sqrt{\frac{τ_2}{\V}}ϕ_2\)+\frac{ϕ_1+τ_1 ϕ_2}{\sqrt{\V τ_2}}-\frac{1}{2} \(\sigma -i\sqrt{\frac{τ_2}{\V}}ϕ_2\)\zeta\,. \label{etaTau}}
Here we also rescaled the  $\varphi_i$'s by the volume  $\V$ of the torus. The Gibbons-Hawking potential is now given by (\ref{V5dRec}) with  
\equ{a_1 =  \frac{1}{\sqrt{\V\,τ_2}}\,, \qquad  a_2 =   \frac{1}{\sqrt{\V τ_2}} \left[ τ_1 +  i \tau_2 \frac{1}{2}\(\frac{1}{ζ}+ζ\)\right] \,. \label{a1a2Tau}}  
Upon Poisson resummation and contour integration, we find
\equ{V=\sqrt{\det {g_{ij}}} \sum_{\vec{n} \in \mathbb Z^{2'}} \frac{1}{\sqrt{n^i n^j g_{ij}}} e^{in^i \varphi_i-| \sigma | \sqrt{n^i n^j g_{ij}} }\,,}
with the metric $g$ on the torus given by
\equ{g_{ij}= \frac{\V}{\tau_2} \left(
\begin{array}{cc}
1 & - \tau_1\\ 
-\tau_1 &  |\tau|^2
\end{array}\right)\label{general metric}\,.}
Finally, the magnetic coordinate is still given by (\ref{etam5d}), with the new definitions (\ref{etaTau}), (\ref{a1a2Tau}), and the replacement $2 a \rightarrow \(\sigma +i\sqrt{\frac{τ_2}{\V}}ϕ_2\) $ in  $\eta_m^{(0)}$.

\subsection{Dyonic Instanton Corrections}
Dyonic instantons are particle-like objects which are the uplift of four-dimensional instantons to five dimensions. Due to the Chern-Simons term  
\equ{\frac{\kappa}{24 \pi^2} A \wedge F\wedge F\,,}
they become electrically charged. Their central charge is given by
\equ{Z_I=\kappa \sigma |n_I| + \frac{|n_I |}{g_{5,0}^2}\,,}
where $g_{5,0}$ is the five-dimensional gauge coupling and $n_I$ is the four-dimensional instanton number. Since these particles are electrically charged, they contribute corrections to the metric preserving the isometry. Hence, their effect is incorporated easily  by replacing $a \rightarrow a+Z_I$ in the definition of the $\O(2)$ multiplet. 

A more interesting contribution to the metric will come from magnetic corrections. These are now given by magnetic strings and incorporating their effect will be studied elsewhere.

\section{Summary and Outlook}\label{Conclusions}
We have derived the expression for a set of Darboux coordinates on a hyperk\"{a}hler manifold, parameterized by $\O(2p)$ projective superfields. Our derivation relies on the projective Legendre transform construction of such manifolds and can be understood as enforcing a consistency condition. The application of our results to the PTN metric leads to the expression for the magnetic coordinate derived by GMN, describing the mutually local corrections to the moduli space metric of $\N=2$ SYM on $\mathbb R^3 \times S^1$. Mutually nonlocal corrections can also be incorporated into the projective Lagrangian, leading to the TBA equation studied by GMN. 

We also applied this method to the study of electric corrections to the moduli space of five-dimensional SYM compactified on $T^2$, providing a projective superspace description of the metric discussed in \cite{Haghighat:2011xx} and the corresponding Darboux coordinates. There are two contributions: an $\O(2)$ part determined by the five-dimensional perturbative prepotential, which reduces to the semiflat part in the 4D limit; and the corrections due to electric particles, which reduce to the instanton corrections of the 4D theory.

There are several  open questions which could be addressed within this formalism. For example, it  could shed new light on the three-dimensional limit  of GMN (recently analyzed in \cite{Chen:2010yr}), corresponding to the Atiyah-Hitchin metric. Regarding the five-dimensional theory, corrections due to magnetic strings could be incorporated in a form analogous to what was done in (\ref{f inst non-local}) for the four-dimensional case, leading to an integral equation for the Darboux coordinates. 

Apart from the Darboux coordinates, another important geometrical object is the hyperholomorphic connection (see for example \cite{HHC}) and it would be interesting  to investigate its description using the $\Upsilon \leftrightarrow \eta$ duality\footnote{The authors wish to thank Greg Moore for this suggestion.}. Finally, it would  be quite interesting if this framework could yield any information about the six-dimensional SYM theory compactified on $T^3$, whose exact moduli space is K3.

\vspace{-20pt}
\section*{\cen{Acknowledgements}}
We gratefully acknowledge Martin Roček for suggesting this problem, sharing with us unpublished notes on which this work was based, and for illuminating discussions and helpful advice throughout the course of this work. We wish to thank Babak Haghighat and Stefan Vandoren for helpful discussions on the 5D theory.  We also thank Greg Moore and Rikard von Unge for encouraging discussions and Sergey Cherkis for useful suggestions about the manuscript. This research work is supported in part by NSF grant no. PHY-0969739.

\appendix
\section{Projectors\label{A:proj}}
Here we give some details on the  Antarctic $\(Π_N\)$ and Arctic $\(Π_R\)$ projectors. They are defined by
\eqs{Π_N=\oint_{C_0} \frac{d \zeta'}{2 \pi i} \frac{1}{\zeta- \zeta'}\,,\qquad Π_R=\oint_{C_0} \frac{d \zeta'}{2 \pi i} \frac{1}{\zeta'- \zeta}\,,}
where $C_0$ is a closed contour enclosing the origin. Consider the Laurent expansion around $\zeta=0$ of the function $f(\zeta)=\sum_{m=-\infty}^{\infty} c_m \zeta^m$. Applying the projector $\Pi_N$, we will need to calculate
\equ{\oint_{C_0} \frac{d \zeta'}{2 \pi i} \frac{\zeta'^m}{\zeta- \zeta'}\,. \nonumber}
Since there's a pole at $\zeta'=\zeta$, we avoid the singularity by moving the pole slightly \textit{outwards} in the radial direction. This can be achieved by introducing the $\epsilon-$prescription
\equ{\oint_{C_0} \frac{d \zeta' }{2 \pi i} \frac{\zeta'^m}{\zeta- \zeta'+\epsilon(\zeta+\zeta')}\,.  \nonumber}
If $m \geq 0$, there are no singularities enclosed by the contour and the integral vanishes. If $m<0$ the residue is simply $\zeta^m$. Thus, only negative powers survive:
\[\oint_{C_0} \frac{d \zeta'}{2 \pi i} \frac{\zeta'^m}{\zeta- \zeta'} =  \left\{ 
\begin{array}{l l}
 0 &  \quad \mbox{if $m \geq 0$ }\\
 \zeta^m &  \quad \mbox{if $m<0$ }\\ \end{array}\,, \right. \]
or
\equ{\oint_{C_0} \frac{d \zeta'}{2 \pi i} \frac{f(\zeta')}{\zeta- \zeta'} = \sum_{m=1}^{\infty} \frac{c_{-m}}{\zeta^m}\,.}
Using the same $ε-$prescription as above, the action of $Π_R$ on $f\(ζ\)$ is given by
\equ{\oint_{C_0} \frac{d \zeta'}{2 \pi i} \frac{f(\zeta')}{\zeta'- \zeta} = \sum_{m=0}^{\infty}c_m\zeta^m\,.}
Thus, $Π_N+Π_R=1$ as expected. In addition to these, we can construct other projectors by using appropriate powers of $ζ/ζ'$. An example of that is $\bar{Π}_N$, which annihilates the non-positive powers of $ζ$. Thus the combinations $Π_N\pm \bar{Π}_N$, annihilate only the $\zeta$-independent term.

\section{\textit{c}-map}\label{cmap}
The \textit{c}-map \cite{CFG:cmap} relates classical hypermultiplet moduli spaces in compactifications of type II strings on a Calabi-Yau threefold to vector multiplet moduli spaces via a further compactification on a circle. In \cite{Gates:1999zv, RVV:cmap}, it was shown that the $c$-map has a natural description in projective superspace. It can be regarded as taking a vector multiplet from four to three dimensions and reinterpreting it as a tensor multiplet when returning to four dimensions. This is possible because in three-dimensions, a vector multiplet is equivalent to a tensor multiplet, which can then be dualized into a hypermultiplet in four dimensions.

This means that given an $\N=2$ holomorphic prepotential $\F\(W\)$ describing a vector multiplet:
\equ{\L_{v}=-\text{Im}\left[∫d^2θ d²ϑ\F\(W\)\right]\,,}
there is a corresponding dual projective hypermultiplet Lagrangian $\G$ describing a hyperkähler moduli space given by
\eqs{\L_{s}=&∫d^2θ d^2\bar{θ}\oint \frac{d \zeta}{2 \pi i \zeta}\G(\zeta; \eta_e)=∫d^2θ d²\bar{θ}∮\frac{dζ}{2 π i ζ}\overline{\text{Im}}\left[\frac{\F(\zeta\eta_e)}{\zeta^2}\right] \nonumber \\
=&-i∫d^2θ d²\bar{θ}∮\frac{dζ}{2 π i ζ}\left[\frac{\F\(\zeta \eta_e\)}{\zeta^2}   - \overline{\F\(\zeta \eta_e\)}  \,\zeta^2\right]\,.}
This expression determines the semiflat projective Lagrangian $f^{sf}$ in (\ref{c-map}).

\references{
\bibitem{Seiberg:1994rs}
N. Seiberg and E. Witten,
``Electric - Magnetic Duality, Monopole Condensation, and Confinement in N=2 Supersymmetric Yang-Mills Theory'',
\npb{426}{1994}{19}; Erratum, \npb{430}{1994}{485} [\arXivid{hep-th/9407087}].

\bibitem{Seiberg:1994aj}
N. Seiberg and E. Witten,
``Monopoles, Duality and Chiral Symmetry Breaking in N=2 Supersymmetric QCD'',
\npb{431}{1994}{84} [\arXivid{hep-th/9408099}].
  
\bibitem{Seiberg:1996nz}
N. Seiberg and E. Witten,
``Gauge Dynamics and Compactification to Three-dimensions'',
In \emph{Saclay 1996, The mathematical beauty of physics}, (1996) 333 [\arXivid{hep-th/9607163}].

\bibitem{KS}
M. Kontsevich and Y. Soibelman,
``Stability Structures, Motivic Donaldson-Thomas Invariants and Cluster Transformations'', 2008, \arXivid{0811.2435} {\color{cyan}\small [math.AG]}.

\bibitem{GMN}
D. Gaiotto, G. W. Moore, A. Neitzke,
``Four-dimensional Wall-crossing via Three-dimensional Field Theory'',
\cmp{299}{2010}{163} [\arXivid{0807.4723} {\color{cyan}\small [hep-th]}].

\bibitem{Alexandrov et al}
S. Alexandrov, B. Pioline, F. Saueressig and S. Vandoren,
``Linear Perturbations of Hyperkähler Metrics'',
\emph{Lett. Math. Phys.} {\bf 87} (2009) 225 [\arXivid{0806.4620} {\color{cyan}\small [hep-th]}];\\
S. Alexandrov, B. Pioline, F. Saueressig and S. Vandoren,
``Linear Perturbations of Quaternionic Metrics'',
\cmp{296}{2010}{353} [\arXivid{0810.1675} {\color{cyan}\small [hep-th]}];\\
S. Alexandrov, B. Pioline, F. Saueressig and S. Vandoren,
``D-Instantons and Twistors'',
\jhep{0903}{2009}{044} [\arXivid{0812.4219} {\color{cyan}\small [hep-th]}];\\
S. Alexandrov,
``D-Instantons and Twistors: Some Exact Results'',
\emph{J. Phys.} {\bf A 42} (2009) 335402 [\arXivid{0902.2761} {\color{cyan}\small [hep-th]}].

\bibitem{Alexandrov:2011va} 
 S. Alexandrov,
``Twistor Approach to String Compactifications: a Review'', 2011, \arXivid{1111.2892} {\color{cyan}\small [hep-th]}.

\bibitem{Haghighat:2011xx}
B. Haghighat and S. Vandoren,
``Five-dimensional Gauge Theory and Compactification on a Torus'',
\jhep{1109}{2011}{060} [\arXivid{1107.2847} {\color{cyan}\small [hep-th]}].

\bibitem{ULMR}
A. Karlhede, U. Lindström and M. Roček,
``Self-interacting Tensor Multiplets in N = 2 Superspace'',
\pl{B}{147}{1984}{297};\\
U. Lindstr\"{o}m and M. Ro\v{c}ek,
``N=2 Super-Yang-Mills Theory In Projective Superspace'',
\cmp{128}{1990}{191}.

\bibitem{Hitchin:1986ea}
N. J. Hitchin, A. Karlhede, U. Lindström and M. Roček,
``Hyperkähler Metrics and Supersymmetry'',
\cmp{108}{1987}{535}.

\bibitem{Lindstrom:2008gs}
U. Lindström, M. Roček,
``Properties of Hyperkähler Manifolds and their Twistor Spaces'',
\cmp{293}{2010}{257} [\arXivid{0807.1366} {\color{cyan}\small [hep-th]}].

\bibitem{Lindstrom:1987ks}
U. Lindström, M. Roček,
``New Hyperkähler Metrics and New Supermultiplets'',
\cmp{115}{1988}{21}. 

\bibitem{Ivanov:1995cy}
I. T. Ivanov, M. Roček,
``Supersymmetric Sigma Models, Twistors, and the Atiyah-Hitchin Metric'',
\cmp{182}{1996}{291} [\arXivid{hep-th/9512075}].

\bibitem{Ooguri:1996me}
H. Ooguri and C. Vafa,
``Summing up D-Instantons'',
\prl{77}{1996}{3296} [\arXivid{hep-th/9608079}].

\bibitem{Seiberg:1996ns}
N. Seiberg, S. H. Shenker,
``Hypermultiplet Moduli Space and String Compactification to Three-dimensions'',
\pl{B}{388}{1996}{521} [\arXivid{hep-th/9608086}].

\bibitem{Gaiotto:2011tf}
D. Gaiotto, G. W. Moore and A. Neitzke,
``Wall-Crossing in Coupled 2d-4d Systems'', 2011, \arXivid{1103.2598} {\color{cyan}\small [hep-th]}.

\bibitem{GonzalezRey:1997qh}
F. Gonzalez-Rey, U. Lindström, M. Roček, R. von Unge, S. Wiles,
``Feynman Rules in N=2 Projective Superspace I: Massless Hypermultiplets'',
\npb{516}{1998}{426} [\arXivid{hep-th/9710250}].

\bibitem{Kuzenko:2009wr}
S. M. Kuzenko,
``N = 2 Supersymmetric Sigma-models and Duality'',
\jhep{1001}{2010}{115} [\arXivid{0910.5771} {\color{cyan}\small [hep-th]}];\\
S. M. Kuzenko,
``Comments on N = 2 Supersymmetric Sigma-models in Projective Superspace'',
\emph{J. Phys.} {\bf A 45} (2012) 095401 [\arXivid{1110.4298} {\color{cyan}\small [hep-th]}].

\bibitem{MR Notes}
Martin Roček, Unpublished Notes.

\bibitem{CFG:cmap}
S. Cecotti, S. Ferrara and L. Girardello,
``Geometry of Type II Superstrings and the Moduli of Superconformal Field Theories'',
\emph{Int. J. Mod. Phys.} {\bf A4} (1989) 2475;\\
S. Ferrara and S. Sabharwal,
``Quaternionic Manifolds for Type II Superstring Vacua of Calabi-Yau Spaces'',
\npb{332}{1990}{317}.

\bibitem{Gates:1999zv} 
S. J. Gates, Jr., T. Hübsch and S. M. Kuzenko,
``CNM Models, Holomorphic Functions and Projective Superspace C-Maps'',
\npb{557}{1999}{443} [\arXivid{hep-th/9902211}].

\bibitem{RVV:cmap}
M. Roček, C. Vafa and S. Vandoren,
``Hypermultiplets and Topological Strings'',
\jhep{0602}{2006}{062} [\arXivid{hep-th/0512206}].

\bibitem{AHcoords}
I. Bakas,
``Remarks on the Atiyah-Hitchin Metric'',
\emph{Fortsch. Phys.} {\bf 48} (2000) 9 [\arXivid{hep-th/9903256}];\\
R. A. Iona\c{s},
``Elliptic Constructions of Hyperkähler Metrics I: The Atiyah-Hitchin Manifold'',
2007, \arXivid{0712.3598} {\color{cyan}\small [math.DG]}.

\bibitem{Lambert:1999ua}
N. D. Lambert and D. Tong,
``Dyonic Instantons in Five-dimensional Gauge Theories'',
\pl{B}{462}{1999}{89} [\arXivid{hep-th/9907014}].

\bibitem{Seiberg:1996bd}
N. Seiberg,
``Five-dimensional SUSY Field Theories, Nontrivial Fixed Points and String Dynamics'',
\pl{B}{388}{1996}{753} [\arXivid{hep-th/9608111}].

\bibitem{Lawrence:1997jr}
A. E. Lawrence and N. Nekrasov,
``Instanton Sums and Five-dimensional Gauge Theories'',
\npb{513}{1998}{239} [\arXivid{hep-th/9706025}].

\bibitem{5D:LSTV}
D. Robles-Llana, F. Saueressig, U. Theis and S. Vandoren,
``Membrane Instantons from Mirror Symmetry'',
\emph{Commun. Num. Theor. Phys.} {\bf 1} (2007) 681 [\arXivid{0707.0838} {\color{cyan}\small [hep-th]}].

\bibitem{Chen:2010yr}
H.-Y. Chen, N. Dorey and K. Petunin,
``Wall Crossing and Instantons in Compactified Gauge Theory'',
\jhep{1006}{2010}{024} [\arXivid{1004.0703} {\color{cyan}\small [hep-th]}].

\bibitem{HHC} 
  A. Neitzke,
  ``On a Hyperholomorphic Line Bundle over the Coulomb Branch'', 2011, [\arXivid{1110.1619}{\color{cyan}\small [hep-th]}];\\
    S. Alexandrov, D. Persson and B. Pioline,
  ``Wall-crossing, Rogers Dilogarithm, and the QK/HK Correspondence'',
  \jhep{1112}{2011}{027} [\arXivid{1110.0466} {\color{cyan}\small [hep-th]}].

}
\end{document}